\documentclass[manuscript]{aastex}
\usepackage{apj}
\shorttitle{A Cepheid is no More}
\shortauthors{Macri {\it et al.}}
\slugcomment{}

\def \vk {V\!-\!K}

\begin{document}

\title{A Cepheid is No More: Hubble's Variable 19 in M33}

\author{L.M.~Macri, D.D.~Sasselov\altaffilmark{1}\&K.Z.~Stanek\altaffilmark{2}}
\affil{Harvard-Smithsonian Center for Astrophysics, 60 Garden St., Cambridge
MA 02138, USA} \email{lmacri, sasselov, kstanek@cfa.harvard.edu}

\altaffiltext{1}{Alfred P. Sloan Foundation Fellow}
\altaffiltext{2}{Hubble Fellow}

\begin{abstract}

We report on the remarkable evolution in the light curve of a variable star
discovered by \citet{hu26} in M33 and classified by him as a Cepheid.  Early
in the 20th century, the variable, designated as V19, exhibited a 54.7-day
period, an intensity-weighted mean $B$ magnitude of $19.59\pm0.23$~mag, and a
$B$ amplitude of 1.1~mag. Its position in the P-L plane was consistent with
the relation derived by Hubble from a total of 35 variables. Modern
observations by the DIRECT project show a dramatic change in the properties of
V19: its mean $B$ magnitude has risen to $19.08\pm0.05$~mag and its $B$
amplitude has decreased to less than $0.1$~mag. V19 does not appear to be a
classical (Population I) Cepheid variable at present, and its nature remains a
mystery. It is not clear how frequent such objects are nor how often they
could be mistaken for classical Cepheids.
\end{abstract}


\section{Introduction}

In his seminal work, ``A spiral nebula as a stellar system, Messier 33,''
\citet{hu26} presented the discovery of 35 Cepheid variables in that galaxy,
which were used to determine its distance. No other searches for variables in
M33 were undertaken for the next fifty years, until the surveys of
\citet{vhk75,sc83, kmw87} that discovered a total of 77 new Cepheids and
recovered many of Hubble's objects.

The DIRECT project \citep{ka98,st98} started in 1996 the first modern CCD-based
search for variables in M33, among other targets. Our first two seasons
surveyed three $10\arcmin\times 10\arcmin$ fields located in the central area
of M33; the variable star content of these fields has been presented in
\citet{ma01a} and \citet{st01}. Since these fields had been covered by all
previous surveys, we cross-correlated our variable star catalog with published
catalogs and successfully identified 57 of the 60 known Cepheids. Two of the
three missing variables were short period Cepheids: B08 from \citet{sc83}
(P=3.2~d) and Q19113 from \citet{kmw87} (P=13.4~d). These variables are faint
and close to our detection threshold; therefore, it was not surprising that
they eluded our detection.

The third missing variable turned out to be a surprising object, and is the
topic of this paper. It is V19 from \citet{hu26}, classified by him as a
54.7-day Cepheid. It is a bright, isolated, and easy to identify star located
some $6\arcmin$ north of the nucleus. Hubble's light curve exhibited a
variation of $1.1$~mag, which is not present in our data.  Additionally, the
mean $B$ magnitude of the object increased by $\sim 0.5$~mag since Hubble's
original observations.

\S2 summarizes the existing photometry for this object; \S3 addresses the
current variability, periodicity, and location of this object in the H-R and PL
diagrams; and \S4 discusses the possible nature of this object and its known
Galactic counterparts.

\section{Photometry}

\subsection{Data from \citet{hu26}}

The discovery of V19 by \citet{hu26} was based on observations obtained at the
Mount Wilson 100-inch telescope in 56 epochs between August 1919 and September
1925, with two additional plates from September 1909 and August 1910.
Magnitudes were measured using the Argelander method, and were based on local
reference stars calibrated through a photometric sequence from the Selected
Area 45. Hubble's original magnitudes suffered from zeropoint and scale
errors, as pointed out by \citet{sn83} and \citet{cs87}.

We calculated a new transformation between Hubble's magnitude scale and the
standard system, using our photometry \citep{ma01b} for Hubble's original
comparison stars. We identified 121 of these stars in our CCD images, using the
finding charts published by \citet{sn83} and performed a least-squares fit
between Hubble's photographic magnitudes and our CCD-based $B$ magnitudes.
We found

\vskip -12pt
\begin{equation}
m_B = 1.62 (\pm0.04) \left [m_{pg}-18.5 \right ] + 19.68(\pm0.02),
\end{equation}

\noindent{where $m_B$ is the $B$ magnitude in the standard system and $m_{pg}$
is the photographic magnitude as measured by Hubble. The relation has a
scatter of 0.18~mag after the rejection of 11 stars which deviated by more
than 0.5~mag. Our coefficients agree with those derived by \citet{cs87}. We
corrected Hubble's original measurements using the newly derived
transformation, and analyzed the data using a suite of programs developed by
the DIRECT project \citep{ka98}. We derived a best-fit period of 54.71 days,
identical to Hubble's original value, and an intensity-weighted mean $B$
magnitude of $19.59\pm0.23$~mag. Figure 1 shows a phased light curve of V19
based on Hubble's recalibrated data. The scatter in the light curve is
consistent with the {\it r.m.s.}  scatter of Equation 1.}

\subsection{DIRECT Data}

We based our identification of V19 on the finding charts of \citet{hu26},
\citet{vhk75} and \citet{hs80}, all of whom mark the same star as the variable.
V19 is the DIRECT object D33J013357.1+304512.4, located at R.A.=1h33m57.1s,
Dec.=+30d45m12.4s (J2000.0). We ruled out a mis-identification of V19 with a
nearby star in Hubble's original finding chart by identifying all variables in
the general vicinity ($< 3\arcmin$). This is a generous search area, given the
fact that all other variables discovered by Hubble were recovered within a few
arcseconds of previously published positions.  The nearest variable of any kind
is D33J013357.1+304455.2, a faint ($V=21.5$) periodic variable located
$17\arcsec$ south of V19. The nearest Cepheid is D33J013354.9+304532.5, with a
period of 36 days and located $39\arcsec$ north-west of V19. The nearest
Cepheid with a period close to Hubble's original period for V19 is
D33J013347.5+304423.2, with a period of 56 days and located $2.5\arcmin$ to the
south-west.

V19 was observed by the DIRECT project a total of 11, 116 and 60 times in
$BVI$, respectively, during the 1996 and 1997 observing seasons. The
description of the reduction and analysis of this data set was presented in
\citet{ma01b}. Additional observations of another DIRECT field which contains
V19 were obtained during the 1999 season; for the purposes of this paper, we
performed a preliminary reduction using the standard DIRECT pipeline
\citep{ka98} and calibrated the resulting photometry using bright, non-variable
stars from the database of \citet{ma01b}. This third year of observations
yielded an additional 19, 121 and 35 points in $BVI$, respectively. The mean
$BVI$ magnitudes of V19 obtained from the three seasons spanned by our data are
$19.08\pm0.05$, $18.21\pm0.05$ and $17.30\pm0.05$~mag, respectively. The light
curves exhibit very little variation, of order 0.05~mag peak-to-peak. \S 3
contains the analysis of our light curve data\footnote{The light curve of V19
and its finding chart can be found at the DIRECT Web page:
\url{http://cfa-www.harvard.edu/\~\/kstanek/DIRECT}}.

\hfill

\subsection{Other photometric data}

We searched the literature for additional observations of V19 obtained after
1926 and before 1996. The only time series photometry that we were able to
locate was that of \citet{vhk75}, who obtained 67 plates of M33 between January
1966 and November 1974 at the Palomar 48-inch Schmidt, using a variety of plate
types and filters. The data for V19 consists of 19 points in $B$ and 41 points
in $V$. Unfortunately, the uncertainties quoted by the authors for the
individual measurements is 0.7~mag, and there are no data available to check
the absolute calibration of their photometric systems. 

V19 was observed by \citet{hs80} in their survey for blue and red stars in M33,
and was designated as R142 in their catalog. We corrected their tabulated
photometry according to \citet{wfm90} and obtained $B=19.1$~mag and $B-V =
0.9$~mag, consistent with our own determinations.  Unfortunately, the authors
did not obtain a spectrum for this star. Additionally, V19 is present in the
fields surveyed by \citet{kmw87}, but these authors do not list it in their
compilation of variables.

\citet{mad85} reported the results of $JHK$ observations of V19 and other M33
Cepheids conducted between 1980 and 1983 at a variety of telescopes using a
photometer with a $5\arcsec$ diameter aperture. However, the authors expressed
concerns about possible contamination by field stars inside their aperture.
Therefore, we analyzed $K_s$-band images of V19, obtained by Don McCarthy with
the PISCES camera \citep{mcc98} at the refurbished MMT, under $0.6\arcsec$
seeing. These images show the presence of another star of similar infrared flux
some $3\arcsec$ away.  A preliminary calibration of the MMT data, based on
isolated 2MASS stars present in the field of view, yields a mean magnitude for
V19 of $K_s=16.1\pm0.05$~mag. This, in turn, implies a $\vk$ color of
$2.1\pm0.1$~mag, which is somewhat redder than, but consistent with, the one
predicted for a 55-day Cepheid, $1.9\pm0.1$~mag \citep[c.f.  Equations 5 and 9
of][]{ma01c}.

Lastly, we retrieved archival HST/WFPC2 frames in which V19 is present (dataset
u2c604, proposal \# 0538) from the Space Telescope Archive. The images were
obtained in the $F439W$ and $F547M$ filters and show the object (at x=92,y=283
on WF4) to be well isolated, with no detectable companions inside our
ground-based seeing disk.

\section{Analysis}

We computed the $J_S$ index \citep{st96} using all 237 $V$ measurements from
the DIRECT data, and found a value of $J_S=0.807$, which is above the standard
DIRECT threshold of 0.75.  However, the overall {\it r.m.s.} scatter is only
0.025~mag, below our usual threshold of 0.04~mag.  Therefore, the star does not
meet the standard criteria for variability adopted by the DIRECT project.

We attempted to find a period from our data, using the standard DIRECT
technique of a modified Lafler-Kinman algorithm \citep{st96} and the CLEAN
algorithm of \citet{rld87}.  No convincing periods were found, which is not
surprising given that our frame-to-frame random photometric error (arising from
the limitations of fixed-position PSF photometry) is of the order of
$0.02-0.03$~mag. We conclude that {\it if} the 54.7-day periodicity seen by
Hubble is still present, its amplitude must be less than 0.05~mag. Long-term
follow-up observations using larger telescopes would be most useful to detect
any small-scale periodicity and/or variability.

Figure 2 shows the $B$-band light curve of V19 from 1910 to 2000, based on the
data from \citet{hu26}, {vhk75}, and this work.  The intensity-weighted mean
magnitude shows a dramatic increase, from $19.59\pm0.23$~mag in 1919-1925 to
$B=19.08\pm0.05$ in 1996-1999, while the amplitude of pulsation decreases from
1.1~mag to undetectable levels. Figure 3 shows the location of V19 in the H-R
diagrams of \citet{ma01b}.  We also plot the location of 130 Cepheids with
$P>10$~days present in the inner region of M33 \citep{ma01a}. Figure 4 shows
the location of V19 in the $BVI$ Cepheid P-L relations defined by the same set
of variables, using Hubble's original period of 54.7 days. These figures
indicate that the modern optical magnitudes of V19 are somewhat brighter than,
but still consistent with, those of 55-day Cepheids in M33.

\section{Discussion}

The cessation of pulsations, or any significant decrease in amplitude (i.e., by
a factor of ten), has never been observed in a classical Cepheid.  V19 may be
the first case, or it may belong to another class of variable stars.  We
present here four classes of variable stars whose pulsational behaviour may
explain the nature of V19.

\vskip 4pt

\noindent{{\em Population II Cepheids}: RU Cam, a 22-day Cepheid which in 1965
abruptly decreased in amplitude from 1~mag to about 0.1-0.2~mag \citep{df66},
was later shown to be a W Virginis variable (Population II Cepheid) that
exhibited a highly unstable and modulated light curve thereafter
\citep{ks93}. Even before the 1965 event, \citet{pg41} had suggested that RU
Cam may be a W Virginis variable.  However, RU Cam is of little relevance to
our case, because in order for V19 to be a Population II Cepheid, it should be
about 2 magnitudes fainter \citep[see][]{al98} than Population I Cepheids of
similar period. Furthermore, W Virginis variables are not known to change their
mean magnitude, although a few RV Tauri variables (longer-period Population II
Cepheids) do that cyclically.}

\vskip 4pt

\noindent {{\em Peculiar Population I Cepheids}: These objects are extremely
rare and entirely confined to much shorter periods. A case in point is the
classical Cepheid V473 Lyr of $P=1.49$ days \citep[e.g.,][]{bu86,an98}.  It has
an amplitude modulation of a factor of 15 which resembles the beating of two
closely spaced pulsation modes. This phenomenon is not understood, but it is
most likely associated with the interaction of high-order modes.  Furthermore,
the interval corresponding to low amplitudes in V473 Lyr has a short duration;
in general, the phenomenon seems to have very little in common with V19. A more
relevant object in this category could be Polaris, now established to be a
Population I Cepheid pulsating in the first overtone mode \citep{fc97}. It has
a period is 3.97 days, and its amplitude has decreased by a factor of 3 over
the past 50 years. Despite earlier reports, this amplitude decrease has
stabilized over the past 15 years; the pulsation amplitude is practically
constant since 1986 \citep{hc00}. As in the case of V473 Lyr, the amplitude
decrease could be due to high-order mode interactions \citep{es01}.  However,
the pulsation amplitude of Polaris has always been very low ($\leq 0.1$~mag in
V), which makes the comparison with V19 difficult.  Finally, we should note
that there are stars inside the Cepheid instability strip that are not
variable, for reasons that are still unknown.}

\vskip 4pt

\noindent{{\em UU Herculis stars}: These are supergiants with Cepheid-like
pulsations that exhibit occasional standstills \citep{sa84}.  Pulsations cease
abruptly and at mid-cycle (corresponding to the mean magnitude of the star),
last a couple of months, and then abruptly start again \citep{sa83}. The UU
Herculis stars may alternate between Cepheid-like and RV Tauri-like pulsations
(over timescales of several years) with periods of 30 to 90 days, which are
always of relatively low amplitude ($\sim$0.3~mag in V). UU Her is the only
exception, with amplitudes as high as 0.6~mag in V. Most stars in this class
have infrared excesses, low metallicities, and show variable emission in the
H$\alpha$ line \citep{kpc97}.}

\vskip 5pt

\noindent{{\em Luminous Blue Variables}: Given its luminosity (M$_{\rm
V}\approx -6.4$~mag), V19 appears to be more closely related to the Luminous
Blue Variables (LBVs), which are not always blue. One example is 164 G Sco
\citep[M$_{\rm V}=-8.4$, T=10,000K, $R\approx 175 R_{\odot}$,
after][]{lba98}. It shows multiperiodic variability ($45-55$ days) of very low
amplitude ($\leq 0.1$~mag in V). Occasionally, LBVs experience dramatic
outbursts and mass loss episodes which could considerably alter their effective
radii and temperatures for extended periods of time. There also exist other
luminous supergiants which are not LBVs but exhibit Cepheid-like pulsations,
such as V810 Cen \citep{ki98}. However, their pulsations have always been
observed to be low ($\leq 0.1$~mag) and multiperiodic.}

\vskip 6pt

Each of the types described above belongs to a different evolutionary state.
The W Virginis (and RV Tauri) variables are most likely low-mass AGB (and
tip-of- or post-AGB) stars (hydrogen- and helium-shell burning). The same
appears to be true for the majority of UU Hers stars (a related object
which "wanders" across the HR diagram is FG Sge). Hence their lower
luminosity and common circumstellar infrared excesses. They often show
emission in the absorption profile of the H$\alpha$ and other strong lines.
There is no evidence that V19 possesses any of the above characteristics and
its luminosity is incompatibly higher. The remaining types of variables (LBVs,
V810 Cen) are very massive supergiants. At a given pulsation period they are
brighter than Population I Cepheids and they are unlikely to be mistaken for
Cepheids, due to their low amplitudes and multi-periodic pulsation.

In summary, V19 does not belong to any of the above types: it is too luminous
to be a low-mass Population II star, and its pulsation was too regular and
high-amplitude 75 years ago to be a massive microvariable supergiant.
Therefore, V19 appears to be a high-mass Population I star that behaves more
like a low-mass Population II RV Tauri variable. 

Given the bright and isolated nature of V19, and its location in the host
galaxy, it is likely that most photographic plates of M33 obtained at 2-m class
(or larger) telescopes during the 20th century will yield precise photometry
for this object. We encourage the astronomical community to contact us if they
are in possession of plate material that might help constrain the past behavior
of V19. Future observations, including spectroscopy and long-term, multi-band
photometric follow-up, will help resolve the nature of this object. Such
resolution is desirable in order to avoid misclassifying objects like V19 as
classical Cepheids in distant galaxies.

\section{Acknowledgments}

We would like to thank Don McCarthy for the PISCES
images of V19 taken at the new MMT. Thanks are due to D. Bersier, N.R. Evans,
J. Kaluzny \& B. Paczynski for helpful comments. LMM would also like to thank
John Huchra for his support and comments. DDS acknowledges support from the
Alfred P. Sloan Foundation and from NSF grant No. AST-9970812. KZS was
supported by Hubble Fellowship grant HF-01124.01 from the Space Telescope
Science Institute, which is operated by the Association of Universities for
Research in Astronomy, Inc., under NASA contract NAS5-26555.

\clearpage

\begin{figure}
\plottwo{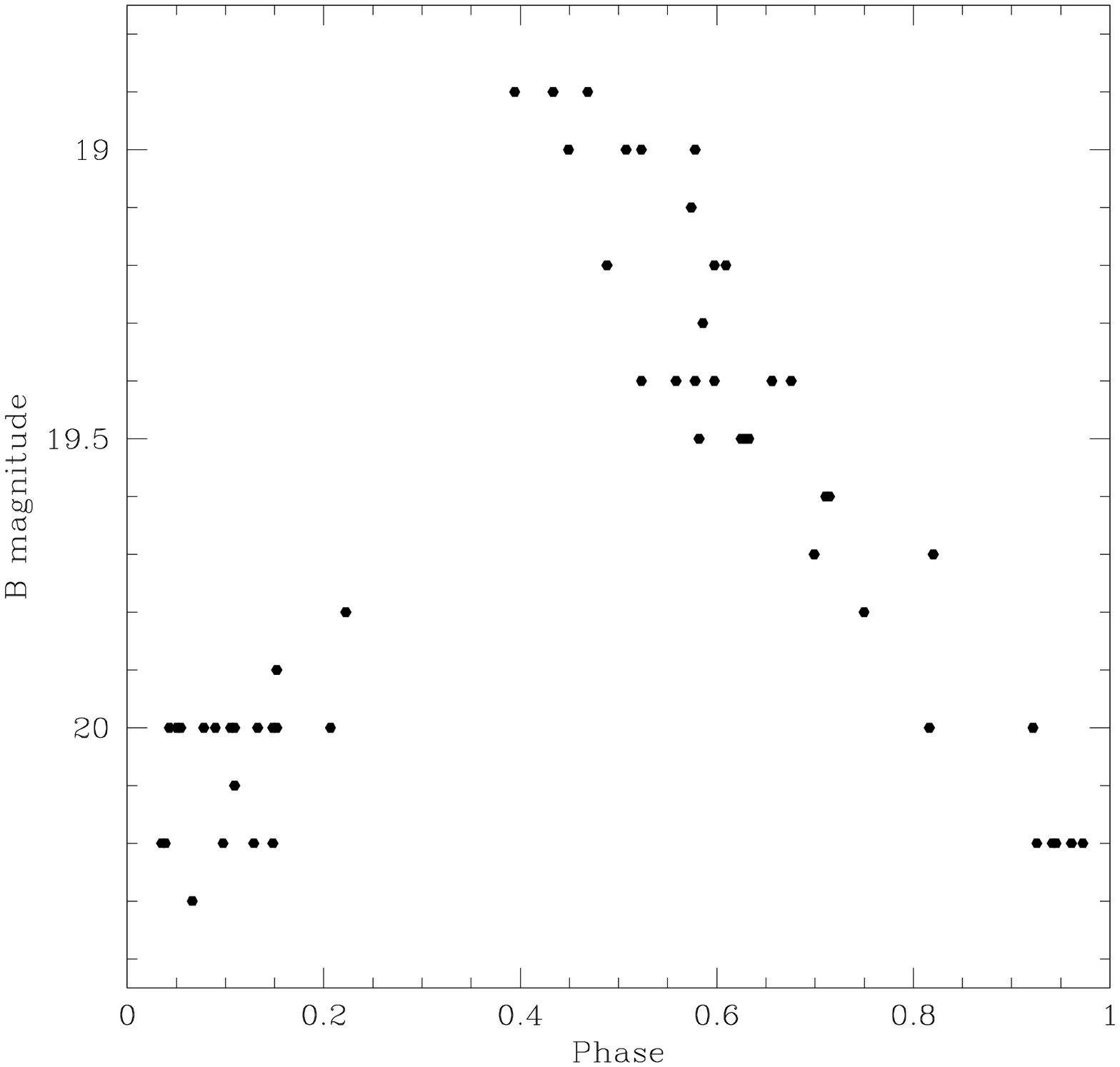}{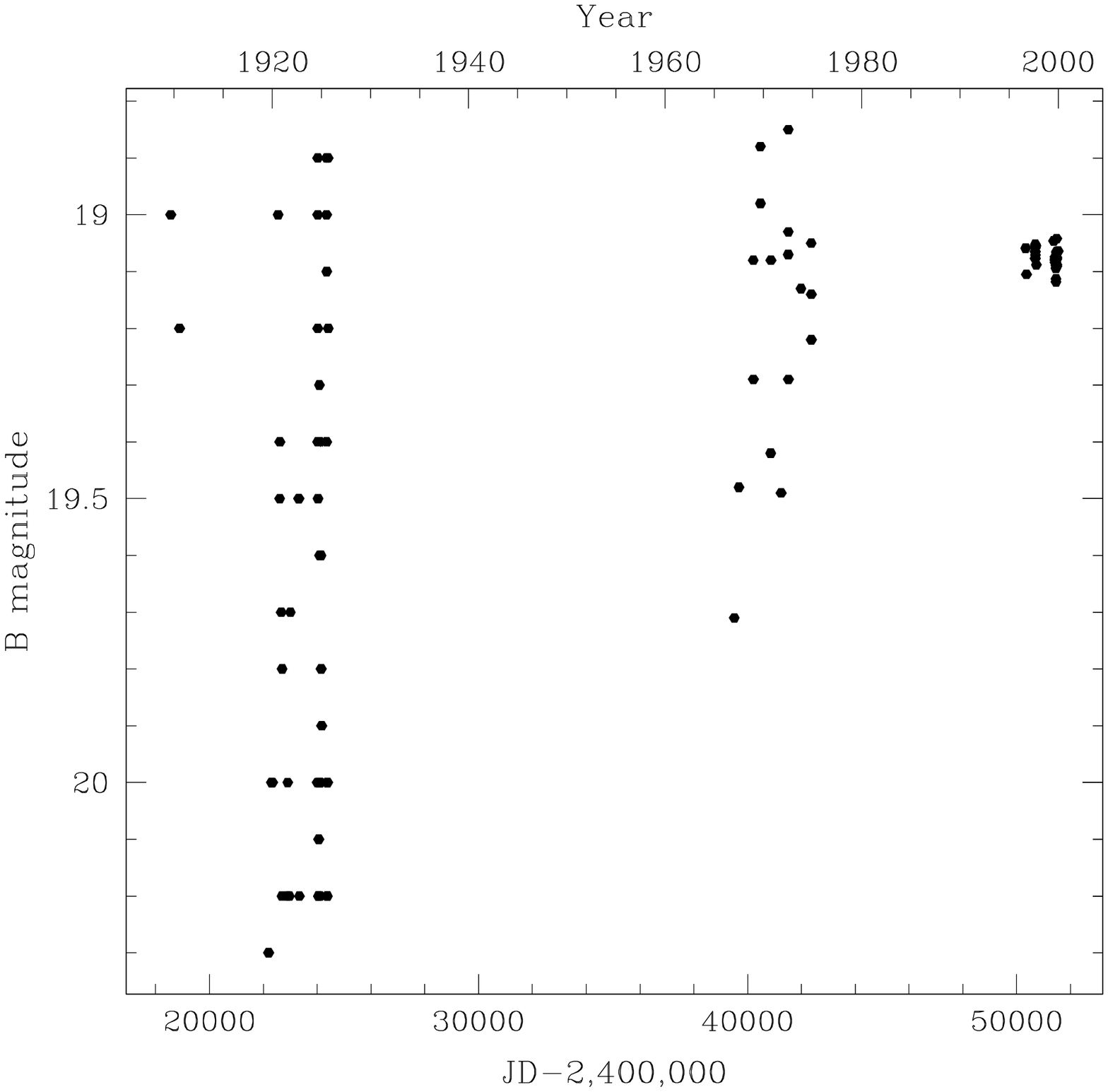}
\caption{({\bf left}) Phased light curve of V19, based on the period and magnitudes published in \citet{hu26}. The original photographic magnitudes have been
corrected into the standard $B$ band using Equation 1.}

\caption{({\bf right}) $B$-band light curve of V19 from 1910 to 1999,
based on the data of \citet{hu26}, \citet{vhk75}, and the present
work.}
\end{figure}

\begin{figure}
\plottwo{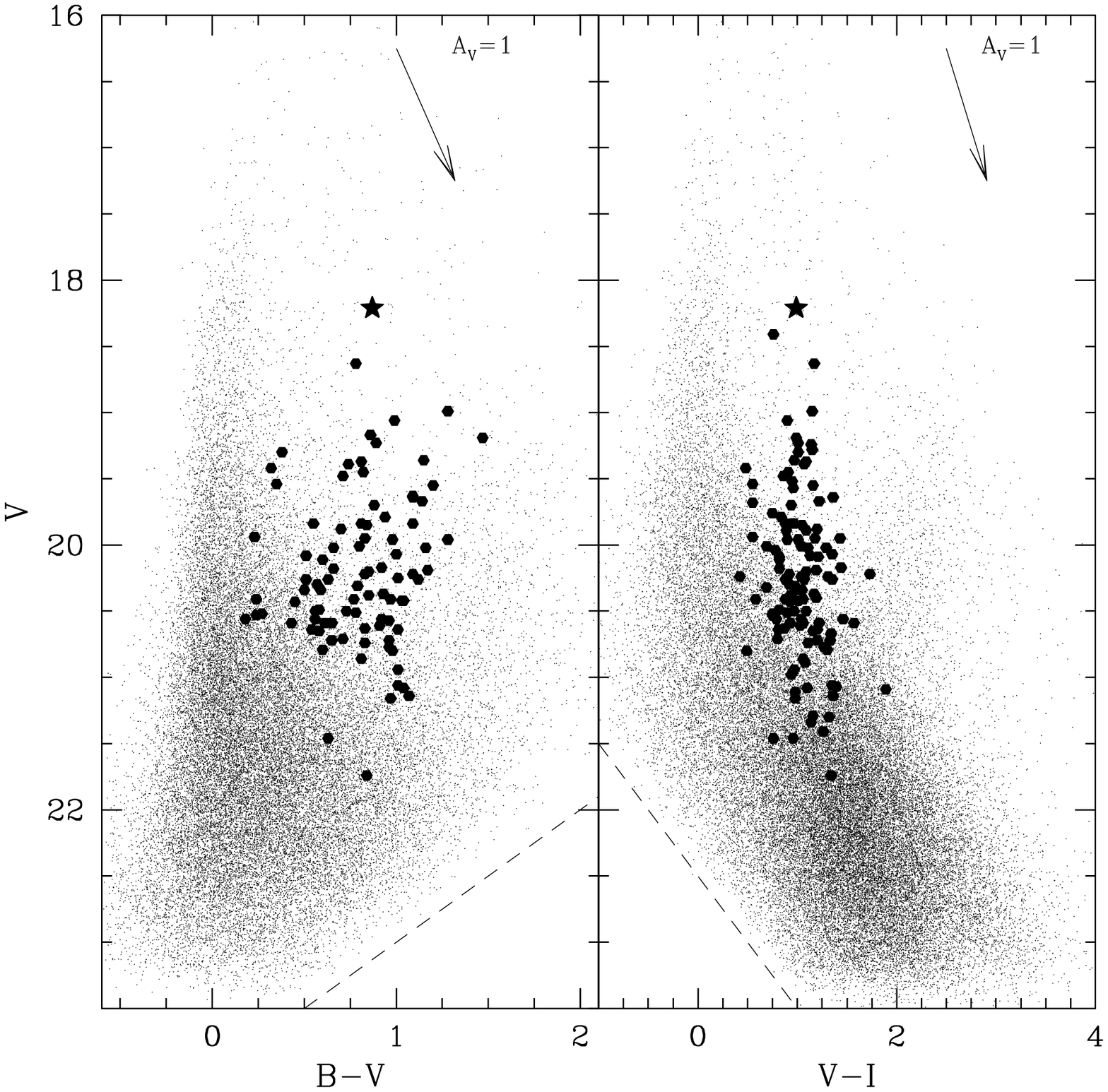}{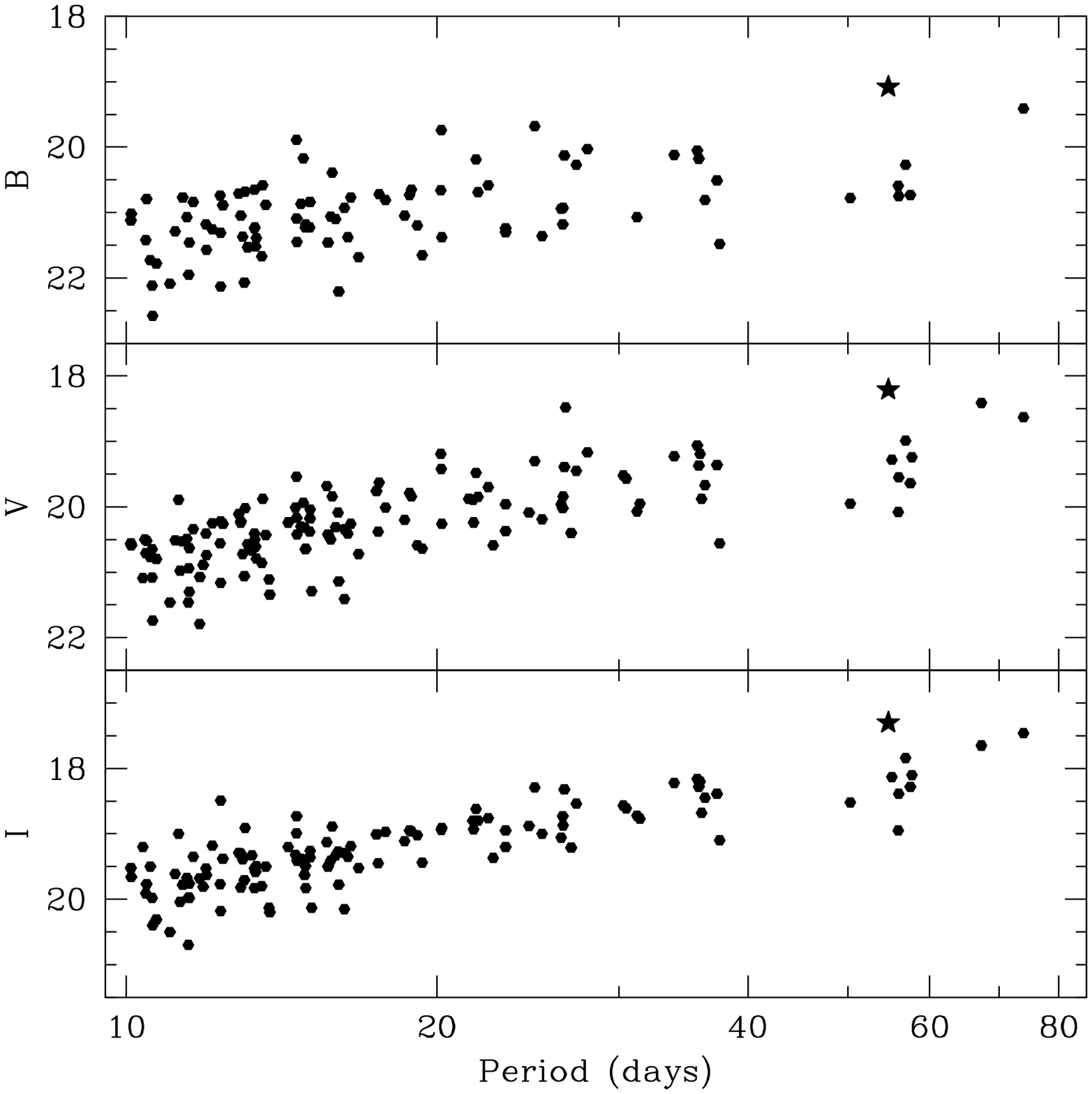}
\caption{({\bf left}) Color-magnitude diagrams of M33 from
\citet{ma01b}, showing the location of V19 ($\star$) and $P>10$~d
Cepheids ($\bullet$).}

\caption{({\bf right)} P-L relations of M33 Cepheids ($\bullet$) from
\citet{ma01a}, showing the location of V19 ($\star$).}
\end{figure}

\begin{thebibliography}{}

\bibitem[Alcock et al.\,(1998)]{al98} Alcock, C. \& MACHO Team,~1998, \aj,
115, 1921

\bibitem[Andrievsky et al.\,(1998)]{an98} Andrievsky S.M., Kovtyukh V.V., 
Bersier, D., Luck, R.E., Gopka, V.P., Yushchenko, A.V., Usenko I.A.~1998,
\aap, 329, 599

\bibitem[Burki et al.\,(1986)]{bu86} Burki G., Schmidt E.G., Arellano Ferro
A., Fernie J.D., Sasselov, D., Simon N.R., Percy J.R., Szabados L.~1986, \aap,
168, 139

\bibitem[Christian \& Schommer\,(1987)]{cs87} Christian, C.A. \& Schommer,
R.A.~1987, \aj, 93, 557 

\bibitem[Demers \& Fernie\,(1966)]{df66} Demers, S., \& Fernie, J.D.~1966,
\apj, 144, 440

\bibitem[Evans et al.\,(2001)]{es01} Evans, N.R., Sasselov, D.D., \& Short,
C.I.~2001, \apj, submitted

\bibitem[Feast \& Catchpole\,(1997)]{fc97} Feast, M.W., \& Catchpole,
R.M.~1997, \mnras, 286, L1

\bibitem[Hatzes \& Cochran\,(2000)]{hc00} Hatzes, A.P., \& Cochran, W.D.~2000,
\aj, 120, 979

\bibitem[Hubble\,(1926)]{hu26} Hubble, E.~1926, \apj, 63, 236

\bibitem[Humphreys \& Sandage\,(1980)]{hs80} Humpreys, R. \& Sandage, A.~1980,
\apjs, 44, 319

\bibitem[Kaluzny et al.\,(1998)]{ka98} Kaluzny, J., Stanek, K.Z.,
Krockenberger, M., Sasselov, D.D., Tonry, J.L. \& Mateo, M.~1998, \aj, 115,
1016

\bibitem[Kienzle et al.\,(1998)]{ki98} Kienzle, F., Burki, G., Burnet, M. \&
Maynet, G.~1998, \aap, 337, 779

\bibitem[Kinman et al.\,(1987)Kinman, Mould \& Wood\,(1987)]{kmw87} Kinman,
T., Mould, J.R., and Wood, P.R.~1987, \aj, 93, 833

\bibitem[Klochkova et al.,\,(1997)]{kpc97} Klochkova, V.G., Panchuk, V.E. \&
Chentsov, E.L.~1997, \aap, 323, 789

\bibitem[Kollath \& Szeidl\,(1993)]{ks93} Kollath, Z., \& Szeidl, B.~1993,
\aap, 277, 62

\bibitem[Lamers et al.\,(1998)]{lba98} Lamers, H., Bastiaanse, M., \& Aerts,
C.~1998, in ASP Conf. Ser. 135, eds. P.Bradley \& J.Guzik, p.159

\bibitem[Macri et al.\,(2001a)]{ma01a} Macri, L.M., Stanek, K.Z., Sasselov,
D.D., Krockenberger, M. \& Kaluzny, J.~2001a, \aj, 121, 861

\bibitem[Macri et al.\,(2001b)]{ma01b} Macri, L.M., Stanek, K.Z., Sasselov,
D.D., Krockenberger, M. \& Kaluzny, J.~2001b, \aj, 121, 870

\bibitem[Macri et al.\,(2001c)]{ma01c} Macri, L.M., Calzetti, D., Freedman,
W.L., Gibson, B.K, Graham, J.A., Huchra, J.P., Hughes, S.M.G., Madore, B.F.,
Mould, J.R., Persson, S.E. \& Stetson, P.B.~2001c, \apj, 549, xxx

\bibitem[Madore et al.\,(1985)]{mad85} Madore, B.F., McAlary, C.W., McLaren,
R.A., Welch, D.L., Neugebauer, G. \& Matthews, K.~1985, \apj, 294, 560

\bibitem[McCarthy et al.\,(1998)]{mcc98} McCarthy, D.W., Ge, J., Hinz, J.L.,
Finn, R.A., Low, F.J., Cheselka, M. \& Salvestrini, K.~1998, \baas, 193, 1109

\bibitem[Payne-Gaposchkin\,(1941)]{pg41} Payne-Gaposchkin, C.~1941, Harvard
Bull., 915, 10

\bibitem[Roberts et al.\,(1987)]{rld87} Roberts, D.H., Lehar, J. \& Dreher,
J.W.~(1987), \aj, 93, 968

\bibitem[Sandage\,(1983)]{sn83} Sandage, A.R.~1983, \aj, 88, 1108

\bibitem[Sandage \& Carlson\,(1983)]{sc83} Sandage, A.R. \& Carlson, G.~1983,
\apjl, 267, 25

\bibitem[Sasselov\,(1983)]{sa83} Sasselov, D.D.~1983, IVBS \# 2387

\bibitem[Sasselov\,(1984)]{sa84} Sasselov, D.D.~1984, \apss, 102, 161

\bibitem[Stanek et al.\,(1998)]{st98} Stanek, K.Z., Kaluzny, J., Krockenberger,
M., Sasselov, D.D., Tonry, J.L. \& Mateo, M.~1998, \aj, 115, 1894

\bibitem[Stanek et al.\,(2001)]{st01} Stanek, K.Z., Macri, L.M., Sasselov,
D.D., \& Kaluzny, J.~2001, in preparation

\bibitem[Stetson\,(1996)]{st96} Stetson, P.B.~1996, \pasp, 108, 851

\bibitem[van den Bergh et al.\,(1975)]{vhk75} van den Bergh, S., Herbst, E. \&
Kowal, C.T.~1975, \apjs, 29, 303

\bibitem[Wilson et al.\,(1990)]{wfm90} Wilson, C.D., Freedman, W.L. \& Madore,
B.F.~1990, \aj, 99, 149

\end{thebibliography}
\end{document}